\documentstyle[version2,preprint,aps]{revtex} 
\renewcommand{\baselinestretch}{1.346}  
\begin{document}
\draft 		
\begin{title}
Competition between point and columnar disorder on the behavior \\
of flux lines in 1 + 1 dimension
\end{title}
\author{T. P. Devereaux, R. T. Scalettar, and G. T. Zimanyi}
\begin{instit}
Department of Physics\\
University of California\\
Davis, CA 95616\\
\end{instit}
\begin{abstract}
We investigate the behavior of flux lines in the presence of both
columnar and point disorder in 1+1 dimension using
Renormalization Group (RG) and world-line
Quantum Monte Carlo (QMC) techniques. In particular, we calculate
the transverse wandering correlation function for a single boson and
recover known results for point and columnar disorder separately.
We then examine the existence of a localization transition of a flux line
in the simultaneous presence of both types of disorder.
We also performed RG for interacting flux lines in the presence of
both disorder.
RG indicates that the vortex glass is unstable with respect to
arbitrary small amount of columnar disorder.
Using QMC we find that the Bose glass transition temperature
is reduced by the point disorder,
in agreement with recent RG calculations. Further we find that
the region posited to be an Anderson glass is completely destroyed.
\end{abstract}
\vspace{0.5 cm}
\narrowtext
\newpage
\section{Introduction}
There has been a great deal of interest lately concerning the
creation of a deliberately disordered environment as a method to pin
flux lines and create a true superconductor with a vanishing linear
resistivity. Two types of disorder have been employed:
point disorder which is uncorrelated in space, and
columnar disorder produced by heavy ion irradiation which is
correlated in a preferred direction\cite{review}.  While
much theoretical attention has been lavished on the
behavior of flux lines in the presence of either point or columnar
disorder and their respective pinned glassy phases, the interplay of the
two types of disorder is still an open topic.
Both types of disorder lead to pinning of flux lines at low temperatures,
but the nature of the phases is quite different.
In the vortex glass\cite{vortex}
the flux line is pinned in a tortuous
meandering path by the point disorder.
In the Bose glass \cite{bose1,bose2}, the flux line
remains relatively straight, locked to the columnar pins.
However the interface of the Vortex and Bose glass phases, where point
and columnar disorder compete, remains beyond
the reach of present analytical approaches\cite{hwa}.
The only guidance comes from a recent study, where
it has been argued that a single flux line will always be localized
for arbitrary weak columnar disorder,
irrespective of the strength of the point disorder\cite{single}.
Interactions between lines could then further limit wandering and
enhance the localizing tendencies.

In this paper we analyze the localization of single and interacting
flux lines in the presence of both point and columnar disorder. In particular
we employ a mapping of the flux line problem in d+1 dimensions
onto the problem of disordered quantum bosons in d dimensions.
We measure transverse flux line wandering of the single flux line,
and discuss the possible existence of
a critical ratio of the columnar to point disorder strength
to localize the boson.  Turning then to the interacting boson problem,
we study the phase boundary of the Bose
glass phase and its stability to point disorder.
We combine these simulations with
a Renormalization Group (RG ) treatment of a disorder potential which
interpolates between point and columnar disordered potentials and find
evidence for trajectories scaling away from the vortex glass phase.
We also find evidence
that a localized phase at weak boson repulsion, the ``Anderson-glass" region,
is destroyed by the de-phasing effect of the point disorder.

\section{formalism}
\subsection{Mapping}
Our Monte Carlo analysis is based on the mapping of flux lines in
d+1 dimensions onto the quantum problem of interacting bosons in
d dimensions.  This mapping can be understood qualitatively
as arising from the similarity between the flux lines aligned
parallel to the applied field and the boson world lines which
appear in a path integral representation.
More specifically, the classical De Gennes - Matricon
free energy in the absence of disorder is\cite{degennes}
\begin{equation}
F_{class}=\int_{0}^{L} dz \left\{\sum_{j=1}^{N}{1\over{2}}\tilde\epsilon_{1}
(dr_{j}(z)/dz)^{2}+
{1\over{2}}\sum_{i\ne j} V(\mid r_{i}(z) - r_{j}(z)\mid)\right\}.
\end{equation}
Here the displacement field $r_{i}(z)$ describes
the position of flux line $i$ in a direction perpendicular
to that of the applied field ($\hat z$).
The interaction potential $V$ between the flux lines is
often simplified to a delta function that
prevents fluxes from occupying the same site. Then one can
define a discrete field that takes on only integer values
\begin{equation}
A(x,z)=\sum_{n}\Theta(x-r_{n}(z)),
\end{equation}
where $\Theta$ is the Heaviside step function, and rewrite the classical free
energy as
\begin{equation}
F_{class}=\int dx \int_{0}^{L} dz \left\{{1\over{2}}\tilde\epsilon_{1}
(dA/dz)^{2}+{V\over{2}} (dA/dx)^{2}\right\}.
\end{equation}
The partition function in the canonical ensemble is then
\begin{equation}
Z_{class}=\int D[A] e^{-F_{class}/T}.
\end{equation}

Consider now the quantum mechanical free energy for bosons in
d spatial dimensions.
In the continuum limit
\begin{equation}
F_{quantum}= \int_{0}^{\hbar\beta} d\tau \int dx \left\{-{t\over{2}}
(\Psi^{\dagger}d^{2}\Psi/dx^{2} + c.c)  +
{U\over{2}} \mid \Psi^{\dagger}\Psi\mid^{2}\right\}
\end{equation}
plus terms which can be absorbed into a chemical potential.
$\Psi$ is the second quantized field representing the bosons.
Following Haldane \cite{haldane}, we can write the Bose density operator
\begin{equation}
\hat N(x)=\Psi^{\dagger}(x)\Psi (x)=
[\partial_{x}\hat B (x)]\sum_{m}e^{-im\hat B(x)},
\end{equation}
where $\partial_{x}\hat B(x)=N_{0} +\hat \Pi(x)$ is a field chosen to
ensure the discrete nature of the Bose field. Here $N_{0}$ is the average
boson density and $\hat\Pi$ is a small fluctuation. Linearizing in $\hat\Pi$
we obtain
\begin{equation}
F_{quantum}= \int_{0}^{\hbar\beta} d\tau \int dx \left\{{\hbar^{2}\over{t}}
(d\hat B/d\tau)^{2} + {U\over{2}} (d\hat B/dx)^{2}\right\},
\end{equation}
where we discard nonlinear terms representing commensuration effects.
Thus we find that the partition function
\begin{equation}
Z_{quantum}=\int D[Q] e^{-F_{quantum}/\hbar}.
\end{equation}
Comparison of the two partition functions leads to the analogy\\

\begin{tabular} {|l | l|} \hline\hline
CLASSICAL FLUXES & QUANTUM BOSONS ($\hbar$)\\
\hline
T & $\hbar$\\
\hline
$L$ & $\hbar$/T\\
\hline
$\tilde\epsilon_{1}/2$ & $\hbar^{2}/t$\\
\hline
V & U\\
\hline
\end{tabular} \\

\noindent
or, if one wants to set $\hbar=1$, then the mapping is\\

\begin{tabular} {|l | l|} \hline\hline
CLASSICAL FLUXES & QUANTUM BOSONS ($\hbar=1$)\\
\hline
$L$ & 1/T\\
\hline
$\tilde\epsilon_{1}/2T$ & $1/t$\\
\hline
V/T & U\\
\hline
\end{tabular} \\

\noindent
and all the energy scales get divided by $T$.
The same holds true when one includes the chemical and disorder
potentials.

\subsection{Simulation Algorithm}

In this paper we consider a discrete lattice version of the interacting boson
problem, the Bose--Hubbard Hamiltonian\cite{bose2}
\begin{equation}
H= -t \sum_{i} (a_{i+1}^{\dagger}a_{i} +
a_{i}^{\dagger}a_{i+1}) +V\sum_{i} n_{i}^{2}+
\sum_{i} n_{i}\epsilon_{c}(i).
\end{equation}
As we shall see, $\epsilon_{c}$ represents columnar disorder.
The inclusion of point disorder will be discussed below.

The partition function $Z$ can be written by discretizing the
imaginary time $\beta=L\Delta \tau$, separating the exponentials of the
kinetic and potential energies, and inserting compete
sets of occupation number states:\cite{SUGAR}
\begin{eqnarray}
Z&=& {\rm Tr} \,\,e^{-\beta H} = {\rm Tr} \,
[e^{-\Delta \tau K} e^{-\Delta \tau P}]^{L}
\nonumber\\
&=&\sum_{n(i,\tau)}
\langle n(i,1) | e^{-\Delta \tau K} e^{-\Delta \tau P} | n(i,2) \rangle
\langle n(i,2) | e^{-\Delta \tau K} e^{-\Delta \tau P} | n(i,3) \rangle
\nonumber\\
\,&.&\,\,\,.\,\,\,.
\langle n(i,L) | e^{-\Delta \tau K} e^{-\Delta \tau P} | n(i,1) \rangle
\end{eqnarray}
The sum is over a classical occupation number field whose first
index runs over the spatial sites in the lattice and whose
second index runs from $1$ to $L$ and labels the ``imaginary time
slice'', that is, the position of
the complete set in the expression for $Z$.
Since $H$ conserves particle number, and $K$ moves
bosons only locally, the allowed configurations
of $n(i,\tau)$ form continuous world--line trajectories which
are the analogs of the flux lines.
The sum over all configurations necessary
to evaluate $Z$ is performed stochastically.

Since $P$ is diagonal in an occupation number representation,
\begin{eqnarray}
Z&=&\sum_{n(i,\tau)} e^{-S_{b}}\,\,
\langle n(i,1) | e^{-\Delta \tau K} | n(i,2) \rangle
\langle n(i,2) | e^{-\Delta \tau K} | n(i,3) \rangle
\nonumber\\
\,&.&\,\,\,.\,\,\,.
\langle n(i,L) | e^{-\Delta \tau K} | n(i,1) \rangle
\nonumber\\
S_{b}&=&V \Delta \tau \sum_{i,\tau}  n(i,\tau)^2 +
 \Delta \tau \sum_{i,\tau}  \epsilon_{c}(i) n(i,\tau)
\end{eqnarray}
The matrix elements of the kinetic energy operator $K$ can be evaluated
using the checkerboard decomposition.\cite{CHECK}
We are now in a position to see how ``point disorder''
can naturally be incorporated as an additional imaginary time dependent
random term in $S_{b}$ which takes the form
\begin{equation}
S_{b}^{p} =  \Delta \tau \sum_{i,\tau}  \epsilon_{p}(i,\tau) n(i,\tau).
\end{equation}

All the standard methods of world line simulations can now be employed.
In addition to Monte Carlo moves which locally distort the world
lines, we also include moves which shift the entire world line
position spatially.  This significantly helps equilibration
in the presence of columnar disorder. The point and columnar disorder
are bounded and uniformly distributed, i.e.,
$-c,-p \le \epsilon_{c,p} \le c,p$,
where $c,p$ stands for columnar and point disorder strengths, respectively.

\section{single flux line}

In this section we consider a single boson in
both a random point and columnar disorder potential and investigate the
localization of the boson as a function of the strength of the respective
disorders, $c/p$. In particular, we focus on the transverse wandering
correlation function,
\begin{equation}
R(\tau)=\langle (x(\tau)-x(0))^{2}\rangle,
\end{equation}
which measures the transverse extension of the boson world line from its
initial position as it traverses the imaginary time axis. It is well known
that in the absence of disorder the boson dynamics is diffusive, with the
diffusion constant governed by $t$, or equivalently, the flux line tension
\cite{nelson} such that for large $\tau$,
\begin{equation}
R(\tau)=t\ \tau, ~~~~~~~~~~~\rm{no\ disorder}.
\end{equation}
Point disorder has been shown to promote flux line wandering due to the
presence of favorable site energies which encourage excursions, and
consequently,
\begin{equation}
R(\tau) \sim \tau^{2\zeta} ~~~~~~~\rm{point\ disorder},
\end{equation}
with the wandering exponent $\zeta=2/3$\cite{kardar}.
Lastly, in the case of columnar disorder simple quantum mechanics leads
to the localization of the flux line for arbitrary pinning and thus
the transverse extension of the flux line is bounded\cite{bose1,bose2}
\begin{equation}
R(\tau) \sim l_{c}~~~~~~~\rm{columnar\ disorder}.
\end{equation}
Since our boson must satisfy periodic boundary conditions,
we study the wandering of the world line at imaginary
time $\tau=\beta/2$ after positioning the boson at
its lowest energy site at time slice $\tau=0$. Our runs use on
average 100 realizations of the disorder.
We first show our results which recover the above limiting behaviors
in Figs. 1 and 2.
For the case of no disorder (Fig. 1) we recover both the exponent
$1$ and the value of the diffusion constant $t$
for the case of 32 site chain, in agreement with \cite{nelson}.
The finite size of the
chain limits the correlation function for large values of hopping $t$ and
for large imaginary time separation $\beta$. For the case of point
disorder (again for a 32 site chain, hopping $t$ set equal to 1), we find that
the wandering exponent $\zeta$ increases
from 1/2 to 2/3 for arbitrary point disorder strength,
which is shown in Fig. 2 for $p=10$ and 25.

Turning on the point disorder $p$ with a fixed value of columnar pinning
strength $c$, we see in Figure 3 that $R$ lifts up at higher $\beta$
reflecting the nature of the point disorder which encourages flux line
wandering. Our behavior
is consistent with that of Kardar\cite{KandZ} who via another method
found numerical evidence, that for $c/p > 0.83$, the flux line is
localized. However, it has been shown that a simple extrapolation
of the localization length to infinity is inadequate and the presence of rare
regions due to the point disorder leads to slow glassy dynamics which keeps
the localization length large but finite\cite{single}. In Fig. 4 we
show the value of the ``localization length,'' i.e., the saturation of the
wandering correlation function $R$, for runs at different
values of $c/p$ which indeed
seems to indicate a crossover to delocalized behavior at a value
consistent with that of Kardar. However, our simulations are limited to
$\beta$ values which cannot provide
a closer analysis near $(c/p)_{critical}$ to reveal whether the localization
length slowly turns over and remains finite at lower values
of $c/p$, as argued in Ref. \cite{single}.

\section{interacting flux lines}

We have seen in the previous section how the interplay between point and
columnar disorder affects the dynamics of flux lines in the absence of
interactions.  We now turn on the
flux line repulsion $V$ to study the various pinned phases of interacting
flux lines. The advantage of using a boson world line approach is that
we can take the on-site interactions into account using the same numerical
techniques as before. However, since the boson world lines are
indistinguishable we found it more useful to measure the superfluid
density $\rho_{s}$ rather than the transverse correlation function
as an indicator of the Bose glass $(\rho_s=0)$ and either the Vortex
glass or Line liquid $(\rho_{s} \ne 0)$ phases. In particular, we
extend the results of our previous study of spatially disordered bosons
by adding point disorder\cite{prl}. It was shown in Ref. \cite{prl}
that there were two localized regions, separated by a
superfluid phase as a function of boson repulsion.
It was posited that the physics of the region at small repulsion
(Anderson Glass) was qualitatively different
than the phase at larger values of the repulsion (Bose Glass).
Therefore the two phases may behave differently as point disorder is
introduced.
Localization of noninteracting bosons in $1D$ is a consequence of the
phase coherent (back) scattering of bosons and therefore the presence
of a time-dependent potential will de-phase the boson
wavefunctions and thus allow the boson to propagate \cite{rts}. However,
in the case of strongly interacting bosons, the boson repulsion
governs localization and it could be that the effects of point
disorder are suppressed.

Recently, a renormalization group (RG) approach has been applied\cite{hwa} to
the problem of flux lines with point and columnar disorder. This technique is
able to capture the Bose and Vortex glass phases as well as the Line liquid
phase which exists at high temperatures. It was shown that the location of the
Bose glass transition is shifted towards higher repulsions (i.e. lower
temperatures) by the point disorder while the Vortex glass phase is unaffected
by the columnar disorder. Also, a phase diagram has been constructed based on
RG, suggesting a Vortex Glass to Bose Glass transition at a critical value of
$c/p$. However, since the RG treatment was
confined to small disorder, this transition could not be accessed in
Ref. \cite{hwa}.

We consider two approaches towards understanding the nature of the Bose and
Vortex glasses in the presence of both point and columnar disorder. First we
will reconsider the RG  approach from Ref. \cite{hwa} where we introduce
a disorder potential which interpolates between columnar and point
disorder potentials. This introduces another small parameter which allows us
to construct the scaling trajectories. Similar ideas were used
by Weinrib and Halperin on a related problem \cite{halperin}. Then we perform
Quantum Monte Carlo simulations to test the results of the RG analysis.

\subsection{Renormalization Group Approach}

We reconsider the coarse-grained free energy in a continuum
description\cite{hwa},
\begin{equation}
{\cal F} = \int dx dz
\bigl\{{U\over{2}}(\partial_{x}u)^{2}+{\tilde\epsilon_{1}\over{2}}(\partial_{z}u)^{2}-V_{0}(x,z)\partial_{x}u - 2V_{0}(x,z) \rho\cos(2\pi[x\rho+u(x,z)])\bigr\},
\end{equation}
where $u$ is the flux displacement-like field, $\rho$ is the average line
density, $U$ is the on-site flux repulsion and
$\tilde \epsilon_{1}$ is the line tension. The disorder potential
$V_{0}$ we take to represent both point and columnar potentials by writing
\begin{equation}
\langle V_{0}(x,z) V_{0}(x^{\prime}, z^{\prime})\rangle \sim
\Delta\delta(x-x^{\prime}) h(\mid z-z^{\prime}\mid),
\end{equation}
where we take $h(z-z^{\prime}) =
{z_{0}^{a}\over{z_{0}^{a}+\mid z-z^{\prime}\mid^{a}}}$, where $z_{0}$ is a
short distance cutoff.
The Fourier transform of $h(z)$ is given by
\begin{equation}
h(k_{z})\sim c_{0}+c_{1} k_{z}^{-(1-a)}
\end{equation}
for small $k_{z}$ and $c_{0,1}$ are positive constants.
Thus we see that if $a > 1$ the potential is governed by short-range
correlated disorder and thus describes point disorder,
while for $a < 1$ the potential is governed by long-range
correlations.
Thus disorder with long-range correlations which decay
at larger distances faster than $1/z$ leads to the same critical
behavior as that due to short-range correlated disorder \cite{halperin}.

Replicating the above free energy and performing the disorder average leads to
\begin{eqnarray}
{\cal F}_{n}/T = \int dx dz \sum_{\alpha,\beta,j} \Bigl\{ {1\over{2}}K_{j}\
\partial_{j} u_{\alpha}\cdot \partial_{j}u_{\beta}~~~~~~~~~~~~~~~~~~~~~~~~~~~
\nonumber \\
-\int dz^{\prime}h(\mid z-z^{\prime}\mid)
\bigl[\Delta_{x}\partial_{x}u_{\alpha}(x,z)\partial_{x}u_{\beta}(x,z+z^{\prime})+g\cos (2\pi[u_{\alpha}(x,z)-u_{\beta}(x,z+z^{\prime})])\bigr]\Bigr\},
\end{eqnarray}
where $\alpha,\beta \in \{1, \cdots n\}$ are replica indices and $j=\{x,z\}$.
The bulk and tilt moduli are given by $K_{x}=U/T$ and
$K_{z}=\tilde\epsilon_{1}/T$, respectively, and the
bare coupling constant of the nonlinear cosine interactions is
$g=\rho^{2}\Delta/T^{2}$. Lastly, $\Delta_{x}=\Delta/T^{2}$.

We have performed a renormalization group analysis of Eq. (20) using
$\delta=1-a$ and $g$ as small parameters. We developed a technique which
is an extension of the sine-Gordon scaling theory as discussed e.g. by
J. Kogut\cite{kogut}. We performed our calculations up to next-to-leading
order in the couplings. This involved considering terms with 8 displacement
fields.  High scaling dimensions eliminate many of the newly generated
terms. As a novelty, in this order the renormalization of the disorder
correlation exponent $a$ became necessary.
Upon rescaling by a factor
$e^{l}$, tedious but straightforward calculations lead to the
recursion relations as
\begin{eqnarray}
{dK_{x}\over{dl}}=0, ~~~~~~~~{dK_{z}\over{dl}}=C_{0}g_{1},~~~~~~~
{dg_{0}\over{dl}}=g_{0}(2-K^{-1}),~~~~~~~~~\nonumber \\
{dg_{1}\over{dl}}=g_{1}(3-a-K^{-1}(1+\Delta_{x}(c_{0}+c_{1})/2K_{x}))-
C_{1}g_{1}^{2}, ~~~~~ {da\over{dl}}=-C_{2}g_{1}^{2}\delta,
\end{eqnarray}
with $g_{0,1}= c_{0,1}g,\  C_{0,1,2}$ are positive constants and
$K^{-1}={2\pi\over{\sqrt{K_{x}K_{z}}}} \sim T$. The
recursion relations recover previous results on the Bose-glass and Vortex-Glass
transitions, namely, for $a=0$
we have the Bose glass transition temperature $K^{-1}_{B.G.}=3$ without
point disorder\cite{giam}. This critical value is decreased by both point
disorder as found in Ref. \cite{hwa}, and by the exponent $a$.  The
Vortex-Glass
transition $K^{-1}_{V.G.}=2$ is unmodified by columnar disorder or the
exponent of the decay of the disorder correlations.
Let us denote the dimensionles measure of disorder by $D$:
$D~=~\Delta_{x}(c_{0}+c_{1})/2K_{x}$.
Now it was argued in Ref.\cite{hwa} that when the point disorder
is strong enough: $D > 1$, then
the point disorder, measured by $g_{0}$,
scales faster to strong coupling than the columnar disorder ($g_{1}$),
indicating a transition to a vortex glass phase. However this conclusion
has to be viewed with some caution.
Strictly speaking when deriving the scaling equations the coefficients
(or equivalently the scaling dimensions)
of {\it both} $g_{0}$ and $g_{1}$ should have been small.
Second, in typical real systems and numerical simulations
the bare values of the forward and backward scattering are expected to be
comparable. While there is no restriction
on $D$ in the theory, $g_0$ has to be small,
restricting the applicability of the results for $D\ll 1$ for typical
systems. Therefore conclusions about a transition around $D\sim 1$
have to be viewed with some caution.

Both of these problems are remedied by the present method.
First we see that for $a\approx 1$ the transition to the vortex glass
phase happens at small disorder: $D = 1-a \ll 1$.
Second, in this region both scaling dimensions are indeed small,
justifying a proper expansion. This formulation thus constitutes
a technically sound description of the transition to the
vortex glass phase.

However let us observe that the disorder exponent $a$ scales in second order:
for $a <1 \ (\delta > 0)$, $a$ scales to zero,
while for $a > 1 \ (\delta <0)$ $a$ scales to large values.
Thus the system which started out on the Vortex Glass side of the transition
$1 > D > 1-a \ll 1$ will scale {\it into} the Bose Glass phase for $a<1$.

This flow into the Bose Glass phase for {\it all} $a<1$ indicates an
instability of the Vortex Glass phase towards the addition of
arbitrarily weak columnar pinning. While the ``weakness" is characterized
by the critical exponent $a$ being arbitrary close to the point-disorder limit
$1$, we expect that this result also translates to a critical value of $c/p
=0$.
This result is also plausible as the same instability has been argued for
single flux lines \cite{single}, and adding interactions will further
localize the vortices.
However, the Bose Glass phase itself is unstable to
point disorder, although admittedly only on astronomically long length scales
\cite{hwa}. The most likely picture emerging from the RG is therefore
that the trajectories flow towards a strong coupling fixed point(s), which
is characterised by simultaneous finite values of $c$ and $p$. Because
of the presence of columnar disorder, this
feature would most likely translate to a finite localization length.

\subsection{Numerical Approach}

To test this scenario we have employed our
Monte Carlo algorithm to investigate the phase boundary between
the insulating and superconducting phases of the boson Hamiltonian,
corresponding to the Bose (large $V$) or Anderson (small $V$) glass
and either the Line liquid or the Vortex Glass phase, respectively.
No attempt was made to distinguish between the Vortex glass and the
Line liquid
phases since it has been difficult to find the Vortex
glass phase from static properties \cite{batrouni}.  Rather, we instead
measure the superfluid density $\rho_{s}$ as a function of both
boson repulsion $V/t=2(\pi K)^{2}$ and $c/p$ to map out
the Bose or Anderson glass phase boundary.

Results of our runs are shown in Fig. 5 for the case of a 64 site chain
with 40 bosons at a temperature
$\beta=16$. Values of $\rho_{s}$
are obtained by averaging over several realizations
of the columnar disorder (fixed at c=2) and point disorder,
although each run showed only slight variations in the superfluid density
and appears to be self averaging. Typical runs were over 50,000 warm-up
and 100,000 measurement sweeps. The bottom curve in Fig. 5 represents
new measurements at lower temperatures of our previous results
for $\rho_{s}$ \cite{prl}. The transition from the
superfluid phase to the localized one at large repulsions occurs at a
slightly greater value than the Giamarchi-Shultz fixed point but
is reduced compared to the runs which were
performed at higher temperatures \cite{prl}.  This may indicate that the
localization length is nearing the chain size at this value of disorder.
Further the localized phase at small repulsion (Anderson glass) becomes
more pronounced at these lower temperatures.

Turning on the point disorder shows that the superfluid phase becomes
enhanced, which is shown in the upper curve in Fig. 5 for $c/p=1$.
The Bose-glass transition now occurs at a larger boson
repulsion, which is in agreement with the RG  results from Eq. (21) and
Ref. \cite{hwa}. Further, the Anderson glass is even more adversely
affected than the Bose glass phase as there is no evidence of a
localized phase at small boson repulsion.  It may be possible that
the localization length is beyond the length of the chain at this
temperature although runs performed at larger $\beta=32$
do not support this.  This would indicate that
the Anderson glass is destroyed due to the de-phasing
effect of the presence of random site energies changing in time,
and is in agreement with previous results concerning Mooij
anomalies \cite{rts}. In addition, we have performed runs at
$c/p=0.1$ and have found no evidence for a localized phase for any
value of boson repulsion.

\section{discussion and conclusions}

In this paper we have presented the analysis of independent and
interacting flux lines in the presence of competing point and
columnar disorder.
Our Quantum Monte Carlo simulations have verified that for
point disorder only the exponent $\zeta$ characterizing line wandering
changes from $1/2$ to $2/3$, in agreement with theoretical predictions.
In the presence of columnar disorder a single flux line
becomes localized. The localization length grows
with the introduction of point disorder. It was argued in Ref. \cite{single}
that the localization length remains finite but very large for large
values of $c/p$.
Our results are not inconsistent with this picture.
We found clear evidence for a finite localization length for $c/p < 0.8$,
but for larger values of $c/p$ the localization length
crosses over to very large values, which are inaccesible with our method.

For interacting flux lines we studied the same issues
with the help of Renormalization Group and Monte Carlo techniques.
We developed a variable exponent description of the disorder,
which allows for a technically sound scaling treatment of the problem.
The scaling trajectories indicate the instability of the Vortex glass phase
for arbitrarily small values of columnar disorder towards a
Bose glass, which is presumably characterized by the localization of
the vortices.
Furthermore as the single flux line appears to be
localized for arbitrarily small columnar
disorder as well, it is likely that adding interactions between the lines will
further enhance localization. Our Monte Carlo technique found a
Bose Glass phase at medium values of the columnar disorder.
The lack of finding this localized phase extending to very small values of $c$
might be due to system size restrictions and the importance of rare regions.
The localization length could thus be large but finite.
It was also shown that the Bose glass phase is unstable towards the addition
of point disorder, but only on astronomically long length scales.
Thus presumably the ultimate asymptotic behavior is governed by
a fixed point with {\it both} c and p finite. As $c>0$, the main physical
characteristic of this phase is expected to be the localization of the
vortex lines.

We continue by noting that previously it has been implicitly assumed
that the relevant quantity which characterises the disordered environment
of flux lines is the ratio of the columnar to point disorder strengths,
$c/p$. However, the recursion relations of Eq. (21) seem to prevent a
reduction in the scaling to a single dimensionless coupling
due to the presence of the second term in the recursion relation
for $g_{1}$, which predicts the reduction of the Bose glass transition
temperature. Further, since the RG of Eq. (21) is limited to small
values of the coupling constants, the second order terms could also lead
to the conclusion that both $c$ and $p$ ($g_{1}$ and $g_{0}$) are relevant
parameters, leading to a three-dimensional phase diagram in $K^{-1}, g_{0}$
and $g_{1}$.  This can be tested numerically and is a subject for
future consideration. Lastly, further numerical simulations are necessary to
to resolve the nature of the strong coupling fixed point(s).

In conclusion, we have studied the behavior of free and interacting
fluxlines in the simultaneous presence of point and columnar disorder.
Renormalization Group analysis indicates in both cases that
the addition of even weak columnar defects localizes the flux lines.
However the physics seems to be described by a strong coupling
fixed point, the nature of which has to be clarified by numerical methods.
To this end Quantum Monte Carlo simulations were also performed, which
show such a localization behavior for medium and strong columnar disorder only.
The possible reason for that may be that the localization length exceeds
the system size in the case of weak columnar pinning.
It was also demonstrated that the transition temperature to the
Bose Glass is reduced by the presence of point disorder.
Further work is necessary
(for instance from dynamical quantities) to distinguish between the Vortex
and line liquid phases.

\acknowledgements

We would like to thank T. Hwa for useful and enlightening discussions.
This work was supported by N.S.F. Grant 92-06023.

\figure{The transverse wandering correlation function for different
positions in imaginary time $\beta$ along the boson world line for the case
of no disorder, Eq. (14). A turnover from linear diffusive behavior
occurs when the flux line begins to see the finite lattice size.}
\centerline{
\setlength{\unitlength}{0.240900pt}
\ifx\plotpoint\undefined\newsavebox{\plotpoint}\fi
\sbox{\plotpoint}{\rule[-0.500pt]{1.000pt}{1.000pt}}%
}
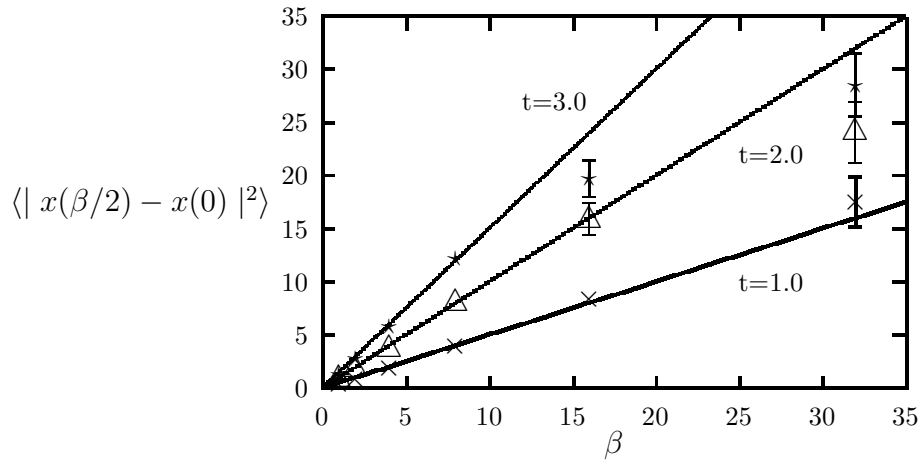
\figure{The transverse wandering correlation function for different positions
in imaginary time $\beta$ along the boson world line for the case
of two point disorder strengths. The slope of the line is $1.33 \pm
0.03$.}
\centerline{
\setlength{\unitlength}{0.240900pt}
\ifx\plotpoint\undefined\newsavebox{\plotpoint}\fi
\sbox{\plotpoint}{\rule[-0.500pt]{1.000pt}{1.000pt}}%
%
}
\newpage
\figure{Transverse wandering correlation function at imaginary time
slice $\beta/2$ for $p=10,c=0$ (crosses); $p=c=0$ (diamonds);\ $p=10, c=5$
(triangles);\ $p=c=10$ ($\times$'s);\ $p=0, c=10$
(squares) for a 32 site chain.}
\centerline{
\setlength{\unitlength}{0.240900pt}
\ifx\plotpoint\undefined\newsavebox{\plotpoint}\fi
\sbox{\plotpoint}{\rule[-0.500pt]{1.000pt}{1.000pt}}%
%
}
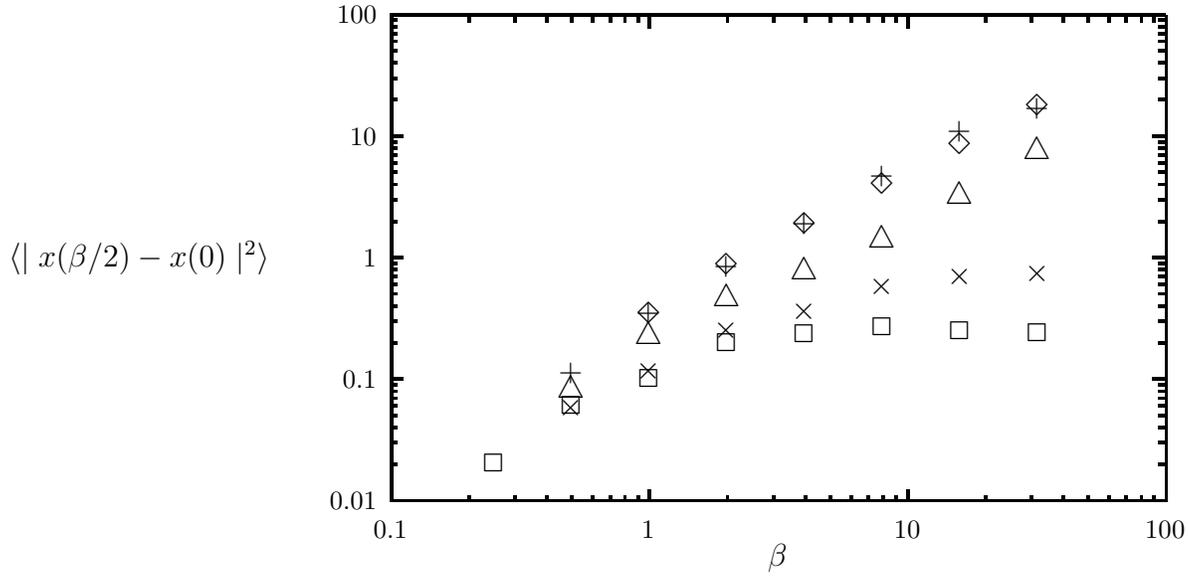
\figure{Square of the inverse localization length as a
function of $p/c$ for a 32 site chain.}
\centerline{
\setlength{\unitlength}{0.240900pt}
\ifx\plotpoint\undefined\newsavebox{\plotpoint}\fi
\sbox{\plotpoint}{\rule[-0.500pt]{1.000pt}{1.000pt}}%
%
}
\newpage
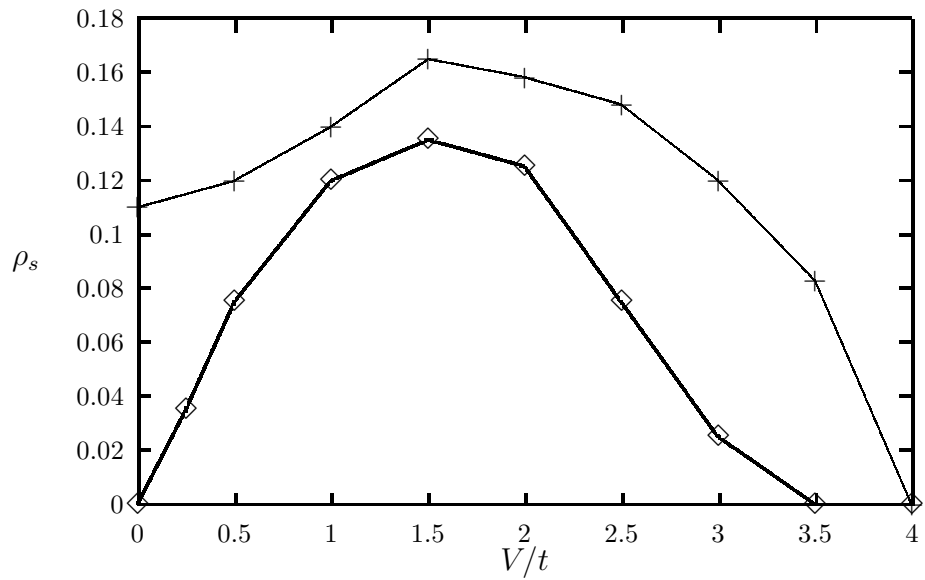
\figure{Superfluid density $\rho_{s}$ for various repulsion strengths
$V/t$ for the case $c=2, p=0$ (bottom line), and $c=p=2$ (top line).}
\centerline{
\setlength{\unitlength}{0.240900pt}
\ifx\plotpoint\undefined\newsavebox{\plotpoint}\fi
\sbox{\plotpoint}{\rule[-0.500pt]{1.000pt}{1.000pt}}%
%
}
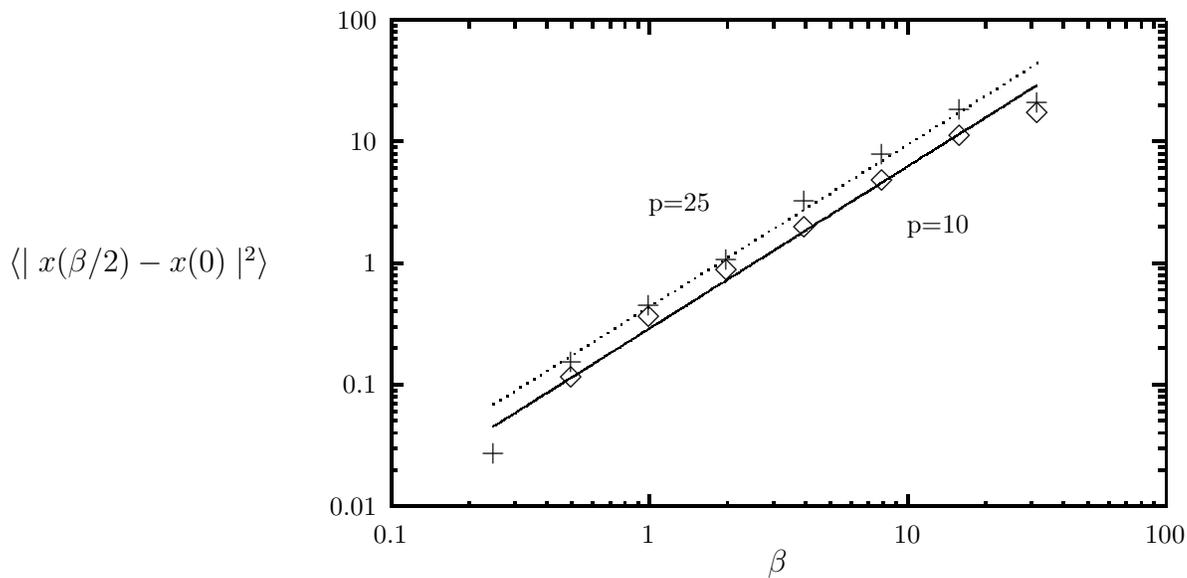
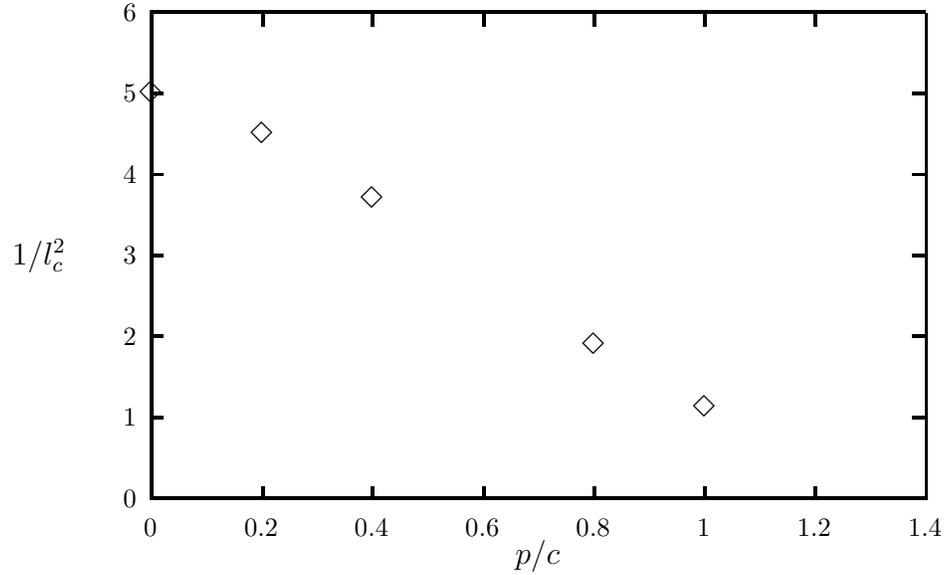
\end{document}